\documentclass[preprint,prd,showpacs,superscriptaddress]{revtex4}
\usepackage{amssymb,amsmath,graphicx,amscd}

\begin{document}

\title{Gravity Waves from Quantum Stress Tensor Fluctuations in Inflation}

\author{Chun-Hsien Wu}
\email{chunwu@scu.edu.tw}
\affiliation{Department of Physics, Soochow University, \\
70 Linhsi Road, Shihlin, Taipei 111, Taiwan}
\author{Jen-Tsung Hsiang}\email{cosmology@gmail.com}
\affiliation{Department of Physics, National Dong Hwa University,
Hualien 97401, Taiwan}
\author{L. H. Ford}\email{ford@cosmos.phy.tufts.edu}
\affiliation{Institute of Cosmology, Department of Physics and
Astronomy,
 Tufts University, Medford, MA 02155, USA}
\author{Kin-Wang Ng}
\email{nkw@phys.sinica.edu.tw}
\affiliation{Institute of Physics,
Academia Sinica, Nankang, Taipei 11529, Taiwan}

\begin{abstract}
We consider the effects of the quantum stress tensor fluctuations of a conformal
field in generating gravity waves in inflationary models. We find a non-scale invariant,
non-Gaussian contribution which depends upon the total expansion factor between
an initial time and the end of inflation. This spectrum of gravity
wave perturbations 
is an illustration of a negative power spectrum, which is possible in
quantum 
field theory. We discuss possible choices for the initial conditions. 
If the initial time is taken to be
sufficiently early, the fluctuating gravity
waves are potentially observable both in the CMB radiation and in gravity wave
detectors, and could offer a probe of transplanckian physics.
 The fact that they have not yet been observed might be used to constrain
the duration and energy scale of inflation. 
\end{abstract}


\maketitle

\baselineskip=14pt

\section{Introduction}

Inflationary models predict a nearly scale invariant  spectrum of both
scalar and tensor perturbations, both of which arise from the quantum
fluctuations of nearly free fields and are Gaussian in character. The
scalar perturbations arise from the the quantum fluctuations of an
inflaton field~\cite{MC81,GP82,Hawking82,Starobinsky82,BST83},
and have apparently been observed in the temperature fluctuations
of the cosmic microwave background~\cite{WMAP}. The tensor perturbations
arise from the fluctuations of quantized linear perturbations of de~Sitter
spacetime~\cite{Starobinsky79,AW84,Allen88}, but have not yet been observed.
In both cases, the Gaussian nature of the fluctuations and the approximate
scale invariance arise from the properties of free quantum fields. 

Coupling of the inflaton or graviton fields to other fields can modify
these conclusions. For example, the coupling of graviton modes to the
expectation value of the quantum stress tensor of a conformal field
was recently treated in Ref.~\cite{HFLY10}. It was shown that
graviton modes can acquire a one-loop correction which increases their
amplitude in a way which depends upon the duration of inflation and
upon the wavenumber of the mode. This effect will tend to lead to a
blue tilt to the spectrum of tensor perturbations, but will not change
their Gaussian character at the one-loop level. 

However, an additional source of perturbations is quantum fluctuations of the 
stress tensor. The effects of stress tensor fluctuations in generating
density perturbations have recently been studied in 
Refs.~\cite{WKF07,FMNWW10}, where a non-Gaussian, non-scale invariant
contribution was found. Furthermore, this contribution can also depend upon
the duration of inflation and  potentially be used to place limits on this
duration. The effect studied in Refs.~\cite{WKF07,FMNWW10} arises from
the quantum fluctuations of the comoving energy density of a conformal field
in its vacuum state. The resulting density perturbations are a
non-Gaussian, non-scale invariant component to be added to the effect
of inflaton field
fluctuations~\cite{MC81,GP82,Hawking82,Starobinsky82,BST83}.

Fluctuations of other components of the stress tensor are capable of
creating tensor perturbations.
The purpose of the present paper is to address the creation of
gravity wave fluctuations by
stress tensor fluctuations of a conformal field in its vacuum state. 
 These can be called passive fluctuations of gravity, as opposed to the
active fluctuations discussed in Refs.~\cite{Starobinsky79,AW84,Allen88}.
The radiation of gravity waves by stress tensor fluctuations of matter
fields in thermal states in flat spacetime was discussed in
Ref.~\cite{DF88}. Matter fields in the vacuum state in flat spacetime
cannot radiate due to energy conservation, but in a time-dependent
 spacetime, such radiation is possible.

Unless otherwise noted, units in which $G = c = \hbar =1$ will be used, where
$G$ is Newton's constant.

\section{Gravitational Radiation in an Expanding Universe}
\label{sec:GREU}

Here we review the formalism needed to compute gravitational
radiation by a time-dependent source. We consider a 
spatially flat Robertson-Walker universe, for which the metric
may be written as
\begin{equation}
ds^2 = -dt^2 +a^2(t)\,  (dx^2 +dy^2 +dz^2)
= a^2(\eta)\,(-d\eta^2 + dx^2 +dy^2 +dz^2)\,.  \label{eq:metric}
\end{equation}
Here $t$ is the comoving time, and $\eta$ the conformal time.
Let $\gamma_{\mu\nu}$ be this background metric, and $h_{\mu\nu}$ be
a linear perturbation,
\begin{equation}
g_{\mu\nu}= \gamma_{\mu\nu} + h_{\mu\nu} \,.
\end{equation}
Here we are concerned with tensor perturbations, and impose the
transverse, tracefree gauge in which
\begin{equation}
h^{\mu\nu}_{~~~;\nu} = 0, ~~~h^\mu_\mu = h = 0, ~~~{\rm and~}~~~ h^{\mu\nu}u_\nu =
0\,,   \label{eq:TT}
\end{equation}
where $u^\nu=\delta^\nu_t$ is the four velocity of the comoving 
observers, and the semicolon denotes the covariant derivative on the
background spacetime.
These conditions remove all of the gauge freedom, and leave the two
degrees of freedom associated with the polarizations of a gravity
wave.

 Lifshitz~\cite{LS} showed that, in the absence of a source, the 
mixed components $ h_\mu^\nu$ 
satisfy the {\it scalar} wave equation,
\begin{equation}
\Box h_\mu ^{~\nu} = 0\,, \label{eq:scalar}
\end{equation}
where 
\begin{equation}
\Box = \frac{1}{\sqrt{-\gamma}} \partial_\mu \left(\sqrt{-\gamma}\,
\gamma^{\mu\nu}\, \partial_\nu \right)
\end{equation}
 is the scalar wave operator for the metric of Eq.~(\ref{eq:metric}).
A consequence of this result is that gravitons in the spatially flat 
Robertson-Walker spacetime behave as a pair of massless,
minimally coupled quantum scalar fields~\cite{FP77}.

In the presence of a source, the metric perturbation satisfies
an inhomogeneous equation
\begin{equation}
\Box h_\mu ^{~\nu} = -16\pi \,S_\mu ^{~\nu} \,, \label{eq:inhomo}
\end{equation}
where $S_\mu ^{~\nu} (x)$ is the transverse, tracefree part of the
stress tensor of the source. It satisfies the conditions in 
Eq.~(\ref{eq:TT}), and is most conveniently defined in momentum
space. 
The solutions of Eq.~(\ref{eq:scalar}) in the spatially flat
Robertson-Walker spacetime may be taken to be plane waves of the form
\begin{equation}
 h_\mu ^{~\nu}(x) =  e_\mu ^{~\nu} \; 
f_k(\eta) \, e^{i {\bf k} \cdot {\bf x}}\,,
\end{equation}
where $ f_k(\eta)$ is a solution of
\begin{equation}
\frac{d}{d\eta} \left( a^2 \, \frac{df}{d\eta} \right) +k^2 a^2\, f =0\,,
\end{equation}
and $ e_\mu ^\nu =  e_\mu ^\nu ({\bf  k},\lambda)$ is a polarization tensor which
satisfies
\begin{equation}
 e_\mu ^{~\mu} =  e_\mu ^{~\nu}\; u_\nu =  e_\mu ^{~\nu}\; k_\nu =0\,.
\end{equation}
If we take vector $ {\bf k}$ to be in the $z$-direction, then the
independent linear polarization tensors can be taken to have the
nonzero components
\begin{equation}
 e_x ^{~x} = -e_y ^{~y} = \frac{1}{\sqrt{2}} \,,
\end{equation}
for the $+$~polarization, or
\begin{equation}
 e_x ^{~y} = e_y ^{~x} = \frac{1}{\sqrt{2}} \,,
\end{equation}
for the $\times$~polarization.

Denote the spatial Fourier transform of any field
$A(\eta,{\bf x})$ by
\begin{equation}
\hat{A}(\eta,{\bf k}) \equiv \frac1{(2 \pi)^3} \int \!\! d^3x \,
e^{i {\bf k} \cdot {\bf x}} A(\eta,{\bf x}) \; .
\end{equation}
In momentum space, the transverse, tracefree part of a stress tensor
is defined by the projection
\begin{equation}
\hat{S}_\mu^{~\nu} (\eta,{\bf k}) = \sum_\lambda   
e_\mu ^{~\beta} ({\bf  k},\lambda)\; 
e_\alpha ^{~\nu} ({\bf  k},\lambda)\, \hat{T}^\alpha_{~\beta}(\eta,{\bf
  k}) \,.   \label{eq:projection}
\end{equation}
Thus given a stress tensor ${T}^\alpha_{~\beta}(\eta,{\bf x})$ in
coordinate space, we first take a Fourier transform to find
$ \hat{T}^\alpha_{~\beta}(\eta,{\bf  k})$, then find 
$\hat{S}_\mu ^{~\nu} (\eta,{\bf k})$ using Eq.~(\ref{eq:projection}),
and finally take an inverse Fourier transform to find 
${S}_\mu^{~\nu} (\eta,{\bf x})$.

Let $G_R(x,x')$ be the retarded
Green's function for the scalar wave operator, which satisfies
\begin{equation}
\Box G_R(x,x')= - \frac{\delta(x-x')}{\sqrt{-\gamma}} \,, \label{eq:GR}
\end{equation}
and $G_R(x,x') = 0$ if $t < t'$. Here $\Box$ is understood to act at
the point $x$. The gravity wave radiated by the source $S_\mu ^\nu$ 
can be written as an integral over the past lightcone of the point $x$
as
\begin{equation}
 h_\mu ^{~\nu}(x) = 16\pi \int d^4x'\sqrt{-\gamma(x')}\,
G_R(x,x')\, S_\mu ^{~\nu}(x') \,. \label{eq:sol}
\end{equation}
The source here could represent either classical matter or quantum fields.
In the latter case, the average effect of a quantum stress tensor can be
described by the semiclassical theory, in which the renormalized expectation
value $\langle T_{\mu\nu} \rangle$ is used as a source in the Einstein equation.

The effects of a conformal quantum field upon graviton modes in de~Sitter
spacetime has recently been treated in the context of the
semiclassical theory~\cite{HFLY10}. It was found that
there is a correction to the graviton modes which grows with increasing duration
of inflation, analogous to the effects found in
Refs.~\cite{WKF07,FMNWW10} and to be discussed in this paper. However,
the effect found in Ref.~\cite{HFLY10} comes only from the the expectation
value of the stress tensor, not from stress tensor fluctuations. If 
$h_\mu ^{~\nu}$ is a classical solution of Eq.~(\ref{eq:scalar}), then
there is a correction term ${h'}_\mu ^{~\nu}$. In the case that the
conformal field is the electromagnetic field, the fractional correction is
\begin{equation}
\Gamma = \left|\frac{{h'}_\mu ^{~\nu}}{h_\mu ^{~\nu}}\right| =
\frac{1}{5\pi} \ell_p^2 \, H\,S\,k \,,
\end{equation}
where $\ell_p$ is the Planck length,
 $H$ is the Hubble parameter during inflation, and $S$ is the
expansion factor during inflation. This effect grows with increasing
$S$ and $k$, but its total magnitude is limited by the requirement
that $\Gamma \alt 1$ for the one-loop approximation to hold.

\section{A Fluctuating Source}
\label{sec:source}

Now we consider the case where the source $ S_\mu ^{~\nu}(x)$ is
undergoing fluctuations, leading to a fluctuating tensor
perturbation, $ h_\mu ^{~\nu}(x)$. The correlation function for
the perturbation is
\begin{equation}
K_{\mu~~\rho}^{~\nu~~\sigma}(x,x')= 
\langle h_\mu ^{~\nu}(x)\; h_\rho^{~\sigma}(x') \rangle 
- \langle h_\mu ^{~\nu}(x)\rangle \langle h_\rho^{~\sigma}(x') \rangle\,,
\end{equation}
and that for the source is
\begin{equation}
C_{\mu~~\rho}^{~\nu~~\sigma}(x,x')= 
\langle S_\mu ^{~\nu}(x)\; S_\rho^{~\sigma}(x') \rangle 
- \langle S_\mu ^{~\nu}(x)\rangle \langle S_\rho^{~\sigma}(x') \rangle\,.
\end{equation}
Their relation follows from Eq.~(\ref{eq:sol}):
\begin{equation}
K_{\mu~~\rho}^{~\nu~~\sigma}(x,x')= 
(16\pi)^2 \int d^4x_1\sqrt{-\gamma(x_1)}\, d^4x_2\sqrt{-\gamma(x_2)}\,
G_R(x,x_1)\,G_R(x',x_2)\, C_{\mu~~\rho}^{~\nu~~\sigma}(x_1,x_2) \,.
\end{equation}
The spatial Fourier transform of this equation may be expressed as
\begin{equation}
{\hat{K}}_{\mu~~\rho}^{~\nu~~\sigma}(\eta,\eta',k) = 64 (2 \pi)^8 \int d\eta_1
d\eta_2 \,a^4(\eta_1)\, a^4(\eta_2)\, \hat{G}(\eta,\eta_1,k)\,
  \hat{G}(\eta',\eta_2,k)\; 
{\hat{C}}_{\mu~~\rho}^{~\nu~~\sigma}(\eta_1,\eta_2,k)\,,  \label{eq:K}
\end{equation}
where ${ \hat{C}}_{\mu~~\rho}^{~\nu~~\sigma}(\eta_1,\eta_2,k)$ and 
$\hat{G}(\eta,\eta',k)$ are the Fourier transforms of 
$C_{\mu~~\rho}^{~\nu~~\sigma} (x_1,x_2) $ and of the retarded Green's
function $G_R(x,x')$, respectively.

If ${\bf k}$ is in the $z$-direction, then the nonzero components
of $\hat{C}_{\mu~~\rho}^{~\nu~~\sigma}(\eta_1,\eta_2,k)$ for the 
$+$~polarization are
\begin{equation}
\hat{C}_+ = \hat{C}_{x~~x}^{~x~~x} =  \hat{C}_{y~~y}^{~y~~y} =
 -\hat{C}_{x~~y}^{~x~~y} = - \hat{C}_{y~~x}^{~y~~x} \,.
\end{equation}
Similarly, the nonzero components for the $\times$~polarization are
\begin{equation}
\hat{C}_\times = \hat{C}_{x~~x}^{~y~~y} =  \hat{C}_{y~~y}^{~x~~x} =
 \hat{C}_{y~~x}^{~x~~y} =  \hat{C}_{x~~y}^{~y~~x} \,.
\end{equation}
In fact, the stress tensor correlation functions for both
polarizations are equal in our case, so we may drop the polarization label
and write
\begin{equation}
\hat{C}(\eta_1,\eta_2,k)= \hat{C}_+(\eta_1,\eta_2,k)=
\hat{C}_\times(\eta_1,\eta_2,k)\,.
\label{eq:C}
\end{equation}

Furthermore, the correlation function $\hat{C}(\eta_1,\eta_2,k)$ 
for the conformal field in
Robertson-Walker spacetime may be related to the corresponding 
correlation function for the conformal field in flat spacetime by a
conformal transformation. First consider a classical stress tensor
$T_\mu^{~\nu}$ in Robertson-Walker spacetime which is conformally
related to ${\cal T}_\mu^{~\nu}$ in Minkowski spacetime. The spatial
components of these tensors are related by  
$T_i^{~j} = a^{-4}{\cal T}_i^{~j}$. The same conformal transformation
applies to the quantum stress tensor correlation function. Although
the conformal anomaly in the expectation value of a quantum stress
tensor operator breaks the conformal symmetry, the conformal anomaly
for free fields is a c-number which cancels in the correlation
function. Consequently, we can write
\begin{equation}
\hat{C}(\eta_1,\eta_2,k)=  a^{-4}(\eta_1)\,  a^{-4}(\eta_2)\,
\hat{C}_{M}(\eta_1-\eta_2,k) \,,  \label{eq:weights}
\end{equation}
where $\hat{C}_{M}(\eta_1-\eta_2,k)$ is the Fourier transform
of the Minkowski spacetime correlation function for a fixed  component
of ${\cal T}_i^{~j}$.  
As is shown in Appendix~\ref{sec:corr_fnts} , it may be expressed as 
\begin{equation}
\hat{C}_{M}(\eta_1-\eta_2,k) = - \frac{k^5}{512 \pi^5}\, \int_0^1 du \, (1-u^2)^2\,
\cos[k u(\eta_1 -\eta_2)] \,.
\label{eq:C-flat}
\end{equation}
This result applies for either polarization.

\section{The Power Spectrum in Inflationary Cosmology}
\label{sec:power}

The well-known  Wiener-Khinchin~\cite{Wiener,Khinchine} theorem 
states that the Fourier
transform of a correlation function is a power spectrum. 
A corollary of this theorem is that the power spectrum can normally
be written as the expectation value of a squared quantity, and
hence must be positive. However, the latter result can fail in 
quantum field theory, and negative power spectra are possible.
This has recently been discussed in Ref.~\cite{HWF10}. Let
$C({t-t',\bf x}-{\bf x'})$ be a flat spacetime correlation function.
We define the associated power spectrum by a spatial Fourier
transform at $t=t'$:
\begin{equation}
P(k) = \frac{1}{(2\pi)^3}\int d^{3}u\, 
{\rm e}^{i\,\mathbf{k}\cdot\mathbf{u}} \,C(0,{\bf u}) \,.
\label{eq:Power_def}
\end{equation}
Given a power spectrum, we can find the correlation function in space
at equal times as an inverse Fourier transform:
\begin{equation}
C(0,{\bf u}) = \int d^{3}k\, 
{\rm e}^{-i\,\mathbf{k}\cdot\mathbf{u}} \, P(k) \,.
\label{eq:inv_Fourier}
\end{equation}
(Here we will use $C$ to denote either a generic or a stress tensor
correlation function, and $K$ to denote metric correlation functions.)

The power spectrum for the gravity wave fluctuations is just a spatial
component of ${\hat{K}}_{\mu~~\rho}^{~\nu~~~\sigma}(\eta,\eta',k)$
in the limit that $\eta' = \eta$, and is the same for both
polarizations. Thus we may combine Eqs.~(\ref{eq:K}) and
(\ref{eq:weights}) to write the power spectrum at conformal time
$\eta =\eta_r$ as
\begin{equation}
P(k) =  64 (2 \pi)^8 \int^{\eta_r}  d\eta_1 d\eta_2 \, \hat{G}(\eta_r,\eta_1,k)\,
  \hat{G}(\eta_r,\eta_2,k)\; \hat{C}_{M}(\eta_1-\eta_2,k)\,.
\label{eq:power1}
\end{equation}
The possible treatments of the lower limits of integration will be
discussed below. 
We should note that the quantity which is usually called the power spectrum
in cosmology is not $P(k) $, but rather
\begin{equation}
 {\cal P}(k) = 4 \pi k^3 \, P(k) \,.
\end{equation}
It is $ {\cal P}(k)$ which is approximately independent of $k$ for the active
gravity wave fluctuations.  The probability distribution for quantum stress tensor
fluctuations is a skewed, hence non-Gaussian distribution with non-zero odd moments,
although the explicit form has only been found in two-dimensional 
spacetime~\cite{FFR10}. Consequently, the gravity wave fluctuations produced by
stress tensor fluctuations will also be non-Gaussian.

The Green's function $\hat{G}(\eta,\eta',k)$ satisfies 
\begin{equation}
\left(\partial_\eta^2 + 2\frac{a'}{a}\, \partial_\eta +k^2 \right)\; \hat{G}(\eta,\eta',k) =
\frac{\delta(\eta-\eta')}{(2\pi)^3\, a^2(\eta') }\,,   \label{eq:G1}
\end{equation}
as may be verified by taking a spatial Fourier transform of
Eq.~(\ref{eq:GR}). Here $a' = da/d\eta$. 
Now we wish to specialize to the case of de~Sitter spacetime, for which the scale factor is
\begin{equation}
a(\eta) = -\frac{1}{H\, \eta}\,
\end{equation}
with $\eta < 0$. 
We may set the scale factor to be unity at the end of inflation, $\eta
= \eta_r$, in which case $\eta_r = -1/H$. 
Now Eq.~(\ref{eq:G1}) becomes
\begin{equation}
\left(\partial_\eta^2 + 2H a \partial_\eta +k^2 \right)\; \hat{G}(\eta,\eta',k) =
\frac{\delta(\eta-\eta')}{(2\pi)^3\, a^2(\eta') }\,.   \label{eq:G2}
\end{equation}
Comparison of this result with Eq.~(71) of Ref.~\cite{WKF07} reveals that $\hat{G}(\eta,\eta',k)$
differs from the Green's function defined in the latter reference by a factor of
$1/[(2\pi)^3\, a^2(\eta')]$. Consequently, we may use the result of Ref.~\cite{WKF07}  to write
\begin{equation}
\hat{G}(\eta,\eta',k) = \frac{H^2}{(2\pi k)^3}\; \left \{ -k(\eta-\eta')\, \cos[k(\eta-\eta')]
+(1+k^2\, \eta \eta')\, \sin[k(\eta-\eta')] \right \} \,.   \label{eq:G3}
\end{equation}
Next we turn to a discussion of some the possible initial conditions which can be
imposed on solutions of Eq.~(\ref{eq:power1}).

\subsection{Sudden Switching}
\label{sec:sudden}

Here we impose the initial condition that the metric fluctuations
vanish at $\eta=\eta_0$. The power spectrum of tensor
fluctuations at the end of inflation, $\eta =\eta_r = -1/H$, is then given by
\begin{equation}
P(k) = P_s(k) =  64 (2 \pi)^8 \int^{-1/H}_{\eta_0}  d\eta_1  \int^{-1/H}_{\eta_0}  d\eta_2 \,
 \hat{G}(\eta,\eta_1,k)\,  \hat{G}(\eta,\eta_2,k)\; \hat{C}_{M}(\eta_1-\eta_2,k)\,.
\label{eq:power2}
\end{equation}
The integrals in Eq.~(\ref{eq:power2}) may be evaluated, using for example the algebraic
computer program {\it Mathematica}. In the limit that $k |\eta_0| \gg 1$, the result is 
approximately
\begin{equation}
P_s(k) \approx -\frac{H^4\, \eta_0^2}{3 \pi^3 \, k} (1 + k^2\, H^{-2})
\,.
\label{eq:power-sudden}
\end{equation}

There are several remarkable features of this result: its negative sign, its blue tilt,
and the fact that it grows with increasing $|\eta_0|$. The
possibility of negative power spectra was discussed in
Ref.~\cite{HWF10}, where it was shown that such spectra arise naturally
in quantum field theory for the fluctuations of quadratic operators,
such as quantum stress tensors. Indeed, the power spectrum associated
with the fluctuations of the transverse, tracefree part of the
electromagnetic stress tensor is given by the $\eta_1 = \eta_2$ limit
of Eq.~(\ref{eq:C-flat}),
\begin{equation}
\hat{C}_{M}(0,k) = - \frac{k^5}{960 \pi^5}\, ,
\end{equation}
which is negative. Negative power spectra are always associated with
coordinate space correlation functions which are singular in the
coincidence limit. This is the case for stress tensor correlation
functions. They are also associated with the opposite correlation
versus anticorrelation behavior as compared with a positive power
spectrum with the same functional form. This means that $C(r)$
changes sign of the sign of $P(k)$ changes, so events that were
correlated become anticorrelated and vice versa.
The spectrum is also not scale invariant,
and tilted toward the blue end of the spectrum because $|{\cal P}_s(k)|$
grows with increasing $k$.

Another feature of Eq.~(\ref{eq:power-sudden}) is that the power
spectrum for the gravity waves grows as $\eta_0^2$, which means that
it is proportional to the square of the scale factor change between the initial
time and the end of inflation. This is analogous to the results found in 
Refs.~\cite{WKF07,FMNWW10} for the power spectrum of density
fluctuations produced by quantum stress tensor fluctuations. In
both cases, the growth of fluctuations can potentially be used to place
upper limits on the duration of inflation, as will be discussed in 
Sec.~\ref{sec:implications}. The net expansion factor during inflation is
\begin{equation}
S = {H}\,{|\eta_0|}\,, \label{eq:S}
\end{equation}
so we may write Eq.~(\ref{eq:power-sudden}) as
\begin{equation}
P_s(k) = -\frac{H^2\, S^2}{3 \pi^3 \, k} (1 + k^2\, H^{-2})
\,.
\label{eq:power-sudden2}
\end{equation}
The coordinate space correlation function associated with this power
spectrum is given by Eq.~(\ref{eq:inv_Fourier}) to be
\begin{equation}
K_s(r) = -\frac{4 H^2\, S^2}{3 \pi^2 \, r^2} \left(1 - \frac{2}{ H^2\,r^2}
  \right)
\,.
\label{eq:corr-sudden}
\end{equation}
This gives the correlation of points at spatial separation $r$ at
equal times. Note that it may be either positive or negative.

Although the power spectrum and the associated correlation function
grow with increasing $S$ or energy scale $k$, the perturbative approach 
used here require that $|K(r)| \ll 1$, which places a limit on the magnitude
of the effect. 

It is informative to compare the results of this subsection with the
flat spacetime limit. If we set $a=1$ in Eq.~(\ref{eq:G1}) and solve
for the flat space Green's function, the result is
\begin{equation}
\hat{G}_M(\eta-\eta',k) = \frac{1}{(2\pi)^3\,k}\; \sin[k(\eta-\eta')]
\,.   
\label{eq:GM}
\end{equation}
If we use this Green's function in Eq.~(\ref{eq:power1}), the
resulting power spectrum becomes
\begin{equation}
P_M(k) = -\frac{5 k}{6 \pi^3} \,,
\label{eq:power-flat}
\end{equation}
where a rapidly oscillating term which depends upon the integration
interval has been dropped. The associated coordinate space metric
correlation function at equal times is
\begin{equation}
K_M(r) = \frac{20 \ell_p^2}{3 \pi^2 \, r^4} \,.
\label{eq:corr-flat}
\end{equation}
This function simply describes Planck-scale fluctuations, which are
presumably unobservable. The main point is that the fluctuations do
not accumulate in flat spacetime due to anticorrelations. In a curved
spacetime, such a de~Sitter space, this is no longer the case, and the
anticorrelated fluctuations need not cancel. In the calculations,
the crucial difference is between the flat spce Green's function, 
Eq.~(\ref{eq:GM}), and that in de~Sitter space, Eq.~(\ref{eq:G3}).

\subsection{Exponential Switching}
\label{sec:exp}

In the previous subsection, the interaction between the quantum stress
tensor and the gravitational field was taken to be switched on suddenly
at $\eta = \eta_0$. One might be concerned that either the sign of
$P(k)$, or its growth with increasing $|\eta_0|$ are artifacts of this
sudden switching. Here we investigate a model in which the interaction
is switched on gradually. We replace the step function $\theta(\eta
-\eta_0)$ by an exponential function, $e^{p \eta}$, with $p >0$. This
function vanishes as $\eta \rightarrow -\infty$, and in the limit of
small $p$, is close to unity by the end of inflation. The effect of
this switching function is effectively to switch on the interaction
on a conformal time scale of order $|\eta_0|$, where $\eta_0 =
-1/p$. Equation~(\ref{eq:power2}) is replaced by
\begin{equation}
P_e(k) =  64 (2 \pi)^8 \int^{-1/H}_{-\infty}  d\eta_1 \, s_e(\eta_1)
\int^{-1/H}_{-\infty}  d\eta_2 \, s_e(\eta_2)\,
 \hat{G}(\eta,\eta_1,k)\,  \hat{G}(\eta,\eta_2,k)\; \hat{C}_{flat}(\eta_1-\eta_2,k)\,,
\label{eq:power2e}
\end{equation}
where the switching function is
\begin{equation}
s_e(\eta) =  e^{p \eta}\,.  \label{eq:switch}
\end{equation}
In the limit of small $p$, Eq.~(\ref{eq:power2e}) leads to
\begin{equation}
P_e(k) \approx -\frac{H^4(1 +k^2/H^2)}{8\pi^2\, k^2\, p} +O(\ln
p) = -\frac{H^3(1 + k^2/H^2)\,S}{8\pi^2\, k^2}\,,
\label{eq:power_exp}
\end{equation}
where $S$, given by Eq.~(\ref{eq:S}), is the expansion between
$\eta = \eta_0 =-1/p$ and the end of inflation.

Again the power spectrum is negative, blue tilted, and grows with increasing
$S$ although now linearly.
In this case, the equal time spatial correlation function is an
inverse Fourier transform of $P_e(k)$ given by
\begin{equation}
K_e(r) = -\frac{ H^3\, S}{4 r} \,.
\label{eq:corr-exp}
\end{equation}
Here we have dropped a delta-function term proportional to 
$\delta({\bf x})$, which will not contribute to measurements made at
distinct spatial locations.
Note that because $a(\eta)= 1/(H|\eta|)$, if the switching time $\Delta
\eta$ is of order $|\eta_0|$, then the scale factor approximately
doubles during the switch-on. For example, $a(\eta/2) = 2 a(\eta)$.
In terms of comoving time $t$, where $a(t) = e^{Ht}$, this
corresponding to a time interval of  $\Delta t \approx 1/H$, or one 
horizon crossing time.  Thus the switch-on time in this model is of
order of the horizon crossing time.

\subsection{Adjustable Width Switching}
\label{sec:adj}

The switching function $s_e(\eta)$ used in the previous subsection
contains only one parameter, $p$, which regulates both the effective
duration of inflation and the period over which the switching occurs.
It is instructive to consider a more general function with two
parameters:
\begin{equation}
s_{aw}(\eta) = \frac{1}{1+ e^{ (\eta_0-\eta)/\alpha}}\,.  \label{eq:aw_switch}
\end{equation} 
This function, analogous to the Fermi-Dirac distribution function,
changes from zero to unity when $\eta \approx \eta_0$
over a time scale of $\Delta \eta \approx \alpha$. The resulting
power spectrum, $P_{aw}(k)$, is given by Eq.~(\ref{eq:power2e}), with
$s_e(\eta)$ replaced by $s_{aw}(\eta)$, and may be expressed as
\begin{equation}
P_{aw}(k) =  -\frac{H^4}{2\pi^3\, k^3}\, \int_0^1 du \, (1-u^2)^2\, (I_C^2 
+I_S^2) \,.
\end{equation}
Here
\begin{equation}
I_C = \int_{x_r}^\infty dx \, g(x,x_r) \, s(x)\, \cos (u x) \,,
\end{equation}
and
\begin{equation}
I_S = \int_{x_r}^\infty dx \, g(x,x_r) \, s(x)\, \sin (u x) \,,
\end{equation}
with
\begin{equation}
 g(x,x_r) = (x-x_r) \cos(x-x_r) -(1+x x_r) \sin(x-x_r)\,.
\end{equation}
We use the notation, $x =-k \eta$, $x_r =-k \eta_r$, and $s(x) =   s_{aw}(\eta)$. 
The dominant contributions to the integrals in $I_C$ and $I_S$ come from values
of $x$ of order $x_0 =-k \eta_0 \gg x_r$, so we may write
\begin{equation}
 g(x,x_r) \approx x( \cos x    -  x_r \sin x) \,.
\end{equation}

The resulting integrals may be evaluated using Eqs.~(\ref{eq:sin-int}) and
(\ref{eq:cos-int}), which are derived  in Appendix~\ref{sec:integrals}. 
The final result, when $\alpha \agt 1/k$, is 
\begin{equation}
P_{aw}(k) \approx -\frac{H^4\, \eta_0^2}{2 \pi \, k^2 \, \alpha} (1 + k^2\, H^{-2})
\,.
\label{eq:power-aw}  
\end{equation}
Apart from numerical factors, this result contains both
Eqs.~(\ref{eq:power-sudden}) and (\ref{eq:power_exp}) as special
cases. If  $\alpha \approx 1/k$, then we return to the sudden
switching case of Eq.~(\ref{eq:power-sudden}). On the other hand,
if $\alpha \approx |\eta_0|$, we find Eq.~(\ref{eq:power_exp}), up to
numerical constants. The fact that the constants do not match exactly
may be due to the approximation used in deriving Eqs.~(\ref{eq:sin-int}) and
(\ref{eq:cos-int}) ($q \ll 1$) not being very good near $u=0$.  In summary,  if 
$\Delta \eta \alt  1/k$, we obtain the sudden result,
Eq.~(\ref{eq:power-sudden2}), proportional to $S^2$, and if 
$\Delta \eta \approx  |\eta_0|$, we obtain
Eq.~(\ref{eq:power_exp}), proportional to $S$. Intermediate switching
times lead to Eq.~(\ref{eq:power-aw}).

\section{Implications of the Power Spectrum}
\label{sec:implications}

\subsection{Initial Conditions and the Transplanckian Issue}
\label{sec:transplanck}

Although the results in the previous section depend somewhat on the rate
at which the coupling between the quantum stress tensor fluctuations and the
gravitational field is switched on, in all cases the power spectrum grows as
a power of $S$, the expansion from the initial time to the end of inflation.
Thus we need an interpretation which suggests a reasonable value for this
time, $|\eta_0|$. One possibility is to take this time to be the onset of inflation.
This imposes the initial condition that the gravity wave perturbations vanish
at the beginning of inflation. In this case, $S$ becomes the total expansion factor 
during inflation.  A possible objection to this interpretation is that it can lead to
contributions from transplanckian modes. This raises the question of whether
our perturbative treatment can be trusted, as relations such as Eq.~(\ref{eq:sol})
are lowest order approximations in the dimensionless coupling constant 
 $(\ell_p k)^2$. The transplanckian issue has been
extensively discussed in the contexts of the Hawking effect and of cosmology.
Hawking's original derivation~\cite{Hawking} of black hole radiance
relies 
upon modes which begin far above the Planck energy. The fact that the Hawking effect gives a
beautiful unification of gravity, thermodynamics, and quantum theory can be 
considered to be a powerful argument to take  transplanckian modes seriously.
It is true that it is possible to derive the Hawking effect without  transplanckian 
modes~\cite{Unruh,CJ96}, but only at the price of introducing modified
dispersion relations which break local Lorentz symmetry and hence postulate
new physics. There has been an extensive discussion of the possible role of
transplanckian modes in inflationary cosmology. (See Ref.~\cite{FMNWW10}
for a lengthy list of references.) The effect discussed in this paper has the
potential to serve as an observational probe of transplanckian physics.

There is an alternative  possibility~\cite{FMNWW10}, which is to take the initial 
time at which the perturbation vanishes to depend upon the mode, and to be
 the time at which a  given mode redshifts below the Planck scale in the comoving 
 frame. This avoids the transplanckian issue, but at the price of introducing a non-local
 and frame dependent prescription, which is analogous to introducing non-Lorentz                
invariant modified dispersion relations. In the remainder of this paper, we will
explore the consequences of adopting the former prescription whereby $S$
is the total expansion factor during inflation. 

The dependence of the gravity wave spectrum upon a positive power of $S$ might
seems to contradict a theorem due to Weinberg~\cite{Weinberg}, which was generalized
by  Chaicherdsukal~\cite{Chai07}. This theorem states that radiative corrections during
inflation should not grow faster than a logarithm of the scale factor. However, as
was discussed in more detail in  Ref.~\cite{FMNWW10}, density
perturbations which are proportional to a power of $S$ are really due to
high frequency modes at the initial time, and are hence always large rather than growing.
This comment also applies to the effects found in Ref.~\cite{HFLY10} and in the present paper.

\subsection{Numerical Estimates}

We may use the coordinate space correlation functions, $K_s(r)$ and $K_e(r)$  ,
to estimate the physical effects of the gravity wave fluctuations on various scales.
However, these functions describe the primordial fluctuations at the end of inflation.
After the end of inflation, modes which are outside the horizon remain
approximately constant until they re-enter the horizon. (For a more
detailed discussion, see, for example, Ref.~\cite{N96}.) 
After that point, they redshift with their
amplitude proportional to $1/a$. Let $a_{Hc}$ be the value of the scale factor at
which a mode associated with coordinate length $r$ re-enters the horizon, and
$a_{now}$ be the present value of the scale factor. The present value of the
correlation function is then
\begin{equation}
K_{now}(r) = K(r)\; \left(\frac{a_{Hc}}{a_{now}}\right)^2 \,.
\label{eq:corr-now}
\end{equation}
For the sudden switch model, this becomes
\begin{equation}
K_{now-S}(r) = -\frac{4 H^2\, S^2\, \ell_p^4}{3 \pi^2 \, r^2} \,
 \left(1-\frac{2}{H^2 r^2} \right)\, \left(\frac{a_{Hc}}{a_{now}}\right)^2,
\end{equation}
and for the exponential switch model it is
\begin{equation}
K_{now-E}(r) = -\frac{ H^3\, S \, \ell_p^4}{4 r} \, \left(\frac{a_{Hc}}{a_{now}}\right)^2\,.
\end{equation}
Here the factors of the Planck length $\ell_p$ are written explicitly.

Let $E_R$ be the reheating energy at the end of inflation. This energy has
since been redshifted to that of the cosmic microwave background. We 
set $a=1$ at the end of inflation, so that
\begin{equation}
a_{now} \approx \frac{E_R}{2.5 \times 10^{-4} eV}\,.
\end{equation}
The proper length scale today associated with coordinate distance $r$ is
 \begin{equation}
\ell = a_{now}\, r \,.
\end{equation}
We assume that reheating is efficient, so the vacuum energy at the end of inflation
is of order $E_R^4$, and
 \begin{equation}
H^2 = \frac{8 \pi}{3}\, \ell_p^2 \, E_R^4\,.
\end{equation}
We also assume that the scale of interest was outside the horizon at the end of 
inflation, so that $H r > 1$.
We may combine all of these results to write
\begin{equation}
|K_{now-S}| = 10^{45} \left(\frac{\ell_p}{\ell}\right)^2 \, \left(\frac{E_R}{10^{16} GeV}\right)^6\, \left(\frac{a_{Hc}}{a_{now}}\right)^2 \; S^2 \,,
\label{eq:C_nowS}
\end{equation}
and 
 \begin{equation}
|K_{now-E}| = 10^{11} \left(\frac{\ell_p}{\ell}\right) \, \left(\frac{E_R}{10^{16} GeV}\right)^7\,
\left(\frac{a_{Hc}}{a_{now}}\right)^2 \; S \,.
\label{eq:C_nowE}
\end{equation}
 
Let us first consider the case of  perturbations of the order of the present horizon
size, $\ell \approx 10^{61}\, \ell_p$. In this case, $a_{Hc} \approx a_{now}$.
Data from the WMAP satellite~\cite{WMAP} constrain these
perturbations to satisfy $h \alt 10^{-5}$, so that $|K_{now}|  \alt 10^{-10}$.
Consequently, the sudden switch model leads to
\begin{equation}
 S \alt  10^{34} \,  \left(\frac{10^{16} GeV}{E_R}\right)^3\,,
 \end{equation}
 and the exponential switch model to
 \begin{equation}
 S \alt  10^{40} \,  \left(\frac{10^{16} GeV}{E_R}\right)^7\,.
 \end{equation}
 These constraints on the total expansion during inflation are compatible with
 adequate inflation to solve the horizon and flatness problems, $S \agt 10^{23}$.
 Because $K < 0$, quantum stress tensor fluctuations during inflation will
 tend to produce anti-correlated gravity wave fluctuations. Note that in this 
 example, $|K_{now}| \approx K(r) \alt 10^{-10}$, so the criterion for the validity
 of the perturbative calculation, $|K(r)| \ll 1$, is satisfied.
 
Now we wish to consider perturbations which are well within the present horizon.
For this purpose, we need an approximate model for the current matter
content 
of the Universe. Although the dominant component today is the dark energy, this is likely
to be a recent phenomenon. If the dark energy is due to a cosmological constant term,
it does not redshift and hence does not grow as we go backwards in time. Here
we assume that the Universe was radiation dominated, $a \propto t^{1/2}$, for
$t \alt t_{eq}$ and subsequently matter dominated,  $a \propto t^{2/3}$. Furthermore,
we assume
 \begin{equation}
\frac{a_{eq}}{a_{now}} \approx 10^{-4}\,,
\end{equation}
 so that $t_{eq}/t_{now} \approx 10^{-6}$. A perturbation with proper length $\ell$ 
enters the horizon at $t=t_{Hc} = \ell$. If we assume that $\ell < t_{eq}$, then we 
may write
\begin{equation}
\left(\frac{a_{Hc}}{a_{now}} \right)^2 \approx 10^{-63} \;\frac{\ell}{\ell_p}\,. 
\end{equation}
If we insert this relation into Eq.~(\ref{eq:C_nowS}), the result is 
\begin{equation}
|K_{now-S}| = 10^{-58}\;  \left(\frac{100 \, km}{\ell} \right)\;
\left(\frac{E_R}{10^{16} GeV}\right)^6 \; S^2 \,.
\label{eq:C_nowS2}
\end{equation}
In the case of the exponential switch model, the factors of $\ell$ cancel,
\begin{equation}
|K_{now-E}| = 10^{-52}\;  \left(\frac{E_R}{10^{16} GeV}\right)^7 \; S \,,
\label{eq:C_nowE2}
\end{equation}
leading to a scale independent correlation function on scales 
$\ell \alt 10^{23} \, cm$.

If the magnitude of these fluctuations is sufficiently large, they should produce
background noise in gravitational wave detectors, which has not been observed.
LIGO has placed limits~\cite{LIGO}
 of $h \alt 10^{-24}$ on scales of the order of $10^2\,km$,
corresponding to $| K_{now}| < 10^{-48}$, and leading to the constraints
\begin{equation}
S < 10^{23} \, \left(\frac{10^{10} GeV}{E_R}\right)^3 \,.
\end{equation}
for the sudden switch model, and
\begin{equation}
S < 10^{25} \, \left(\frac{10^{13} GeV}{E_R}\right)^7 \,,
\end{equation}
for the exponential switch model.
However, these results are  compatible with adequate inflation to solve the horizon
and flatness problems only if
\begin{equation}
E_R \alt 10^{10}\, GeV 
\label{eq:ER_constraint}
\end{equation}
for the sudden switch model, and
\begin{equation}
E_R \alt 10^{13}\, GeV 
\label{eq:ER_constraint2}
\end{equation}
for the exponential switch model. In this example, 
$|K(r)| \approx 10^{23} K_{now} \alt 10^{-25}$, so again the requirement
that $|K|$ be small is fulfilled.

\section{Summary and Discussion}
\label{sec:sum}

We have seen that quantum stress tensor fluctuations are capable of creating
gravity waves during inflation. The resulting spectrum has several properties,
including negative power and an amplitude which grows with increasing duration
of the inflationary period. Negative power spectra, although forbidden by the
Wiener-Khinchin theorem~\cite{Wiener,Khinchine}, can arise in quantum
field theory~\cite{HWF10}, especially in quantum stress tensor fluctuations.
A negative power spectrum can be viewed as interchanging correlations and
anti-correlations, as compared to a positive power spectrum of the same
functional form. 

We find that the amplitude of the gravity wave spectrum is proportional to 
a positive power of $S$, the change in scale factor during inflation. A similar 
dependence was also found in Refs.~\cite{WKF07,FMNWW10}, for the effects of
stress tensor fluctuations on density perturbation and in Ref.~\cite{HFLY10}
for the correction to gravity wave modes from expectation value of the stress
tensor of a conformal field. The gravity wave power
depends somewhat upon the details of the initial conditions, being $S^2$ if
one integrates the equations directly from a state of zero fluctuations, and
being $S$ if the interaction between the fluctuating matter stress tensor is
supposed to be switched on over a finite interval of the order of the horizon size
in comoving time.   In all cases, the primordial 
power spectrum of  gravity wave fluctuations
is negative and greater in magnitude at shorter wavelengths. This non-scale
invariant spectrum of fluctuations will be highly non-Gaussian,
due to the non-Gaussian character of quantum stress tensor fluctuations.

These gravity wave fluctuations are potentially observable. Longer wavelengths
could alter the polarization of the CMB, and be detected in the same way as
the active fluctuations. Shorter wavelengths could potentially be detected by
Earth or space based gravity wave detectors. The fact that they have not yet been
detected can be used to infer constraints on the duration and energy scale of
inflation.

\begin{acknowledgments}
We would like to thank Shun-Pei Miao, Richard Woodard,
and the participants of the 14th and 15th Peyresq workshops for valuable
discussions. 
 This work is partially supported by the National Science Council, Taiwan, ROC 
 under the Grants NSC 98-2112-M-001-009-MY3 and NSC99-2112-M-031-002-MY3,
  and by the U.S. National Science Foundation under Grant PHY-0855360.  
\end{acknowledgments}

\appendix
\section{Flat Space Stress Tensor Correlation Functions}
\label{sec:corr_fnts}

In this appendix, we derive the explicit expressions for the flat space correlation
functions utilized in Sect.~\ref{sec:source}, especially Eq.~(\ref{eq:C-flat}). 
All of the expressions in this appendix refer to flat spacetime, so here we drop the
subscript "M".   We may use the results of Ref.~\cite{FW01}, where the electromagnetic
field stress tensor correlation function was shown to be
\begin{eqnarray}
C^{\mu\nu\sigma\lambda}(x,x') &=&
4\, (\partial_{\mu }\partial_{\nu }D)\, 
(\partial_{\sigma }\partial _{\lambda }D)\, 
+ 2\, g_{\mu \nu }\, (\partial_{\sigma }\partial_{\alpha }D)\, 
(\partial_{\lambda}\partial^{\alpha }D)\,
+ 2\, g_{\sigma \lambda}\, (\partial_{\mu }\partial_{\alpha }D)\, 
(\partial_{\nu}\partial^{\alpha }D)\,                 \nonumber \\
&-& 2\, g_{\mu \sigma }\, (\partial_{\nu }\partial_{\alpha }D)\, 
(\partial_{\lambda}\partial^{\alpha }D)\,
- 2\, g_{\nu \sigma }\, (\partial_{\mu }\partial_{\alpha }D)\, 
(\partial_{\lambda}\partial^{\alpha }D)\,              \nonumber \\
&-& 2\, g_{\nu \lambda }\, (\partial_{\mu }\partial_{\alpha }D)\, 
(\partial_{\sigma }\partial^{\alpha }D)\,
- 2\, g_{\mu \lambda }\, (\partial_{\nu }\partial_{\alpha }D)\, 
(\partial_{\sigma }\partial^{\alpha }D)\,            \nonumber \\
&+&(g_{\mu \sigma }g_{\nu \lambda }+g_{\nu \sigma }g_{\mu \lambda }-
 g_{\mu \nu }g_{\sigma \lambda })\,(\partial_{\rho }\partial_{\alpha }D)\, 
(\partial^{\rho}\partial^{\alpha }D) \,. \label{eq:CV}
\end{eqnarray}
Here
\begin{equation}
D = D(x-x') = \frac{1}{4 \pi^2 (x-x')^2} \label{eq:scalar_2pt}
\end{equation}
is the Hadamard (symmetric two-point) function for the massless scalar field.
For our purposes, it is sufficient to compute a single component, such as
$C^{xyxy}$. The result is
\begin{equation}
C^{xyxy}(\tau,r) = \frac{3}{\pi^2[(t-t')^2 - r^2]^4} \,,
\end{equation}
where $r = |\mathbf{x}-\mathbf{x'}|$ and $\tau = t - t'$. 
The spatial Fourier transform of this expression is
\begin{equation}
\hat{C}^{xyxy}(\tau,k) = -\frac{1}{512 \pi^5}\, \left( \frac{d^4}{d\tau^4}
+2k^2 \, \frac{d^2}{d\tau^2} +k^4 \right) \; \left( \frac{\sin k \tau}{\tau} \right)\,, 
\end{equation}
 or equivalently,
\begin{equation}
\hat{C}^{xyxy}(\tau,k) = - \frac{k^5}{512 \pi^5}\, \int_0^1 du \, (1-u^2)^2\,
\cos(k u \tau) \,,
\end{equation}   
which is  Eq.~(\ref{eq:C-flat}).    This expression may be verified by checking that
\begin{equation}
C^{xyxy}(\tau,r) = \int d^3k \, 
e^{-i \mathbf{k}\cdot (\mathbf{x}-\mathbf{x'}) }\; \hat{C}^{xyxy}(\tau,k) \,.
\end{equation}

\section{Evaluation of Sampling Function Integrals}
\label{sec:integrals}

In this appendix, we will evaluate some of the integrals needed in Sect.~\ref{sec:adj},
which involve the function $s_{aw}$, defined in Eq.~(\ref{eq:aw_switch}). We begin
with expression 3.411.23 in Ref.~\cite{GR}, which states that
\begin{equation}
\int_{-\infty}^\infty \frac{x\, e^{\mu x}}{1+e^x}\; dx = -\pi^2 \csc(\pi \mu) \, \cot(\pi \mu) \,,
\end{equation}
for $0< Re(\mu) <1$. This implies that
 \begin{equation}
\int_{-\infty}^\infty \frac{e^{\mu x}}{1+e^x}\; dx = \pi \csc(\pi \mu) \,.
\label{eq:mu}
\end{equation}    
This may be verified by taking  a derivative of Eq.~(\ref{eq:mu}) with respect to $\mu$,
and by noting that when $\mu = 1/2$,   this relation becomes
\begin{equation}
\int_{-\infty}^\infty \frac{e^{ x/2}}{1+e^x}\; dx = 2 \int_{0}^\infty \frac{1}{1+y^2}\; dy
= \pi \,.
\end{equation}   
This confirms that there is no additional constant in   Eq.~(\ref{eq:mu}) . Next we 
may take the limit in which $\mu \rightarrow i q$ to write
\begin{equation}
\int_{-\infty}^\infty \frac{e^{i qx}}{1+e^x}\; dx = -\frac{2\pi i}{e^{\pi q} -  e^{-\pi q}}\,.
\label{eq:q}
\end{equation}       
   
However, we need integrals over a semi-infinite range of the form  
\begin{equation}
\int_{x_r}^\infty \frac{e^{i qx}}{1+e^{(x-x_0)/b  }}  \; dx =
b \,e^{iq x_0} \left[  \int_{-\infty}^\infty \frac{e^{i q b z}}{1+e^z}\; dz
- \int_{-\infty}^\frac{x_r-x_0}{b} \frac{e^{i q b z}}{1+e^z}\; dz \right]\,.
\end{equation}
The second integral on the right-hand side of the above equation may be 
approximated by setting the denominator of the integrand to unity:
\begin{equation}
\int_{-\infty}^\frac{x_r-x_0}{b} \frac{e^{i q b z}}{1+e^z}\; dz 
\approx \frac{i}{q\,b} \, e^{i q (x_r -x_0)} +O(e^{-x_0/b}) \,.
\end{equation}
Thus,
\begin{equation}
\int_{x_r}^\infty \frac{e^{i qx}}{1+e^{(x-x_0)/b  }}  \; dx \approx
 -\frac{2\pi i\, e^{i qx_0} }{e^{\pi q} -  e^{-\pi q}}
+ \frac{i}{q} \, e^{i q x_r } \,.
\end{equation}
If we take a derivative with respect to $q$, then the real and
imaginary parts of the resulting expression become, for $q \,b \gg 1$,
\begin{equation}
\int_{x_r}^\infty \frac{x\, \sin q x}{1+e^{(x-x_0)/b  }}  \; dx
\approx
-2 \pi x_0\, b \,\cos (q x_0) \; e^{-qb}
\label{eq:sin-int}
\end{equation}
and  
\begin{equation}
\int_{x_r}^\infty \frac{x\, \cos q x}{1+e^{(x-x_0)/b  }}  \; dx
\approx
2 \pi x_0\, b\, \sin (q x_0) \; e^{-qb} \,.
\label{eq:cos-int}
\end{equation}


\begin{thebibliography}{99}


\bibitem{MC81} V. Mukhanov and G. Chibisov, JETP Lett. {\bf 33}, 532 (1981).

\bibitem{GP82} A.H. Guth and S.-Y. Pi, Phys. Rev. Lett. {\bf 49}, 1110 (1982).

\bibitem{Hawking82} S.W. Hawking, Phys. Lett. B {\bf 115}, 295 (1982).

\bibitem{Starobinsky82}  A.A. Starobinsky,
Phys. Lett. B {\bf 117}, 175 (1982).

\bibitem{BST83}  J.M. Bardeen, P.J. Steinhardt, and
M.S. Turner, Phys. Rev. D {\bf 28}, 679 (1983).

\bibitem{WMAP} E. Komatsu {\it et al.}, Astrophys. J. Suppl. {\bf
192}, 18 (2011),  arXiv:1001.4538.

\bibitem{Starobinsky79} A.A. Starobinsky, JETP Lett. {\bf 30}, 682 (1979).

\bibitem{AW84} L.F. Abbott and M.B. Wise, Nucl. Phys. B {\bf 244}, 541 (1984).

\bibitem{Allen88} B. Allen, Phys. Rev. D {\bf 37}, 2078 (1988).

\bibitem{HFLY10} J.-T. Hsiang, L.H. Ford, D.-S. Lee, and  H.-L. Yu, Phys. Rev. D,
{\bf 83}, 084027 (2011),  arXiv:1012.1582.

\bibitem{WKF07} C.-H. Wu, K.-W. Ng, and L.H. Ford,  Phys. Rev. D
{\bf 75}, 103502 (2007), arXiv:gr-qc/0608002.

\bibitem{FMNWW10} L.H. Ford, S.-P. Miao, K.-W. Ng, R.P. Woodard, and C.-H. Wu,
 Phys. Rev. D {\bf 82}, 043501 (2010), arXiv:1005.4530.

\bibitem{DF88} S. Delcampo and  L.H. Ford,  Phys. Rev. D {\bf 38},
  3657 (1988).

\bibitem{LS} E. M. Lifshitz, Zh. Eksp. Teor. Phys. {\bf 16} 587
  (1946).

\bibitem{FP77}  L.H. Ford and L. Parker, Phys. Rev. D {\bf 16}, 1601 (1977).

\bibitem{Wiener} N. Wiener, Acta. Math, Stockholm {\bf 55}, 117 (1930).

\bibitem{Khinchine} A. Khinchin, Math. Ann. {\bf 109}, 604 (1934).

\bibitem{HWF10} J.-T. Hsiang, C.-H. Wu, and L. H. Ford, arXiv:1012.3226.

\bibitem{FFR10}  C.J. Fewster, L.H. Ford, and T.A. Roman,  Phys. Rev. D {\bf 81},
121901(R) (2010),  arXiv:1004.0179.

\bibitem{Hawking} S.W. Hawking, Commun. Math. Phys. {\bf 43}, 199 (1975).

\bibitem{Unruh} W.G. Unruh,  Phys. Rev. D {\bf 51}, 2827 (1995), gr-qc/9409008.

\bibitem{CJ96} S. Corley and  T. Jacobson,  Phys. Rev. D {\bf 54}, 1568 (1996),
hep-th/9601073.

\bibitem{Weinberg} S. Weinberg, Phys. Rev. D {\bf 72}, 043514 (2005);
{\bf 74}, 023508 (2006).

\bibitem{Chai07} K. Chaicherdsukal, Phys. Rev. D {\bf 75}, 063522 (2007).

\bibitem{N96} K.-W. Ng, Int. J. Mod. Phys. A  {\bf 11}, 3175 (1996).

\bibitem{LIGO} P.J. Sutton, J. Phys. Conf. Ser. {\bf 110}, 062024 (2008).

\bibitem{FW01} L.H. Ford and C.-H. Wu, Int. J. Theor. Phys. {\bf 42}, 15 (2003),
gr-qc/0102063. 

\bibitem{GR} I.S. Gradshteyn and I.M. Ryzhik, {\it Tables of
  Integrals, Series, and Products, 6th Ed.}  
 (Academic Press, New York, 2000).


\end{thebibliography}
\end{document}